\documentclass[12pt]{article}
\usepackage{graphicx}
\title{\bf Radio emission from Colling-Wind Binaries: Observations and Models}
\author{
S.~M.~Dougherty $^1$, J.M.~Pittard$^2$, E.P.~O'Connor$^{1,3}$\\
\vspace{1cm}\\ 
\normalsize $^1$National Research Council, Herzberg Institute of Astrophysics, DRAO, Penticton, Canada \\
\normalsize $^2$School of Physics and Astonomy, U. Leeds, Leeds, UK.\\ 
\normalsize $^3$Physics Dept., U. Prince Edward Island, Charlottetown, PEI., Canada }
\date{\mbox{}}
\newcommand\apj{{ApJ}}
\newcommand\aap{{A\&A}}%
\newcommand\mnras{{MNRAS}}%
\begin{document}
\maketitle
\pagestyle{empty}
%
%
\def\bull{\vrule height .9ex width .8ex depth -.1ex}
\makeatletter
\def\ps@plain{\let\@mkboth\gobbletwo
\def\@oddhead{}\def\@oddfoot{\hfil\tiny\bull\quad
``Massive Stars and High-Energy Emission in OB Associations''; JENAM
2005, held in Li\`ege\ (Belgium)\quad\bull}%
\def\@evenhead{}\let\@evenfoot\@oddfoot}
\makeatother
%
%
\def\beginrefer{\section*{References}%
\begin{quotation}\mbox{}\par}
\def\refer#1\par{{\setlength{\parindent}{-\leftmargin}\indent#1\par}}
\def\endrefer{\end{quotation}}
%
%
{\noindent\small{\bf Abstract:} We have developed radiative transfer
models of the radio emission from colliding-wind binaries (CWB) based
on a hydrodynamical treatment of the wind-collision region (WCR).  The
archetype of CWB systems is the 7.9-yr period binary WR140, which
exhibits dramatic variations at radio wavelengths. High-resolution
radio observations of WR140 permit a determination of several system
parameters, particularly orbit inclination and distance, that are
essential for any models of this system. A model fit to data at
orbital phase 0.9 is shown, and some short comings of our model
described.}
%
%
\section{What are colliding-wind binaries?}
Observations of WR+O binary systems such as WR140 (Williams et
al. 1990; White \& Becker, Dougherty et al. 2005), WR146 (Dougherty et
al. 1996, 2000; O'Connor et al., these proceedings) and WR147
(Williams et al. 1997) reveal synchrotron emission arising from
relativistic electrons accelerated where the massive stellar winds of
the binary companions collide - the WCR. There is strong evidence that
all WR stars (Dougherty \& Williams 2000), and now many O stars (van
Loo et al., 2005, and references therein), that exhibit non-thermal
emission are binary systems.  It is widely accepted that the electrons
are accelerated by diffusive shock acceleration (DSA), and CWB systems
present a unique laboratory for investigating particle acceleration
since they provide higher mass, radiation and magnetic field energy
densities than in supernova remnants, which have been widely used for
such work.

\section{Models of radio emission in CWBs}
\begin{figure}
 \centering
 \includegraphics[width=8.cm,angle=0]{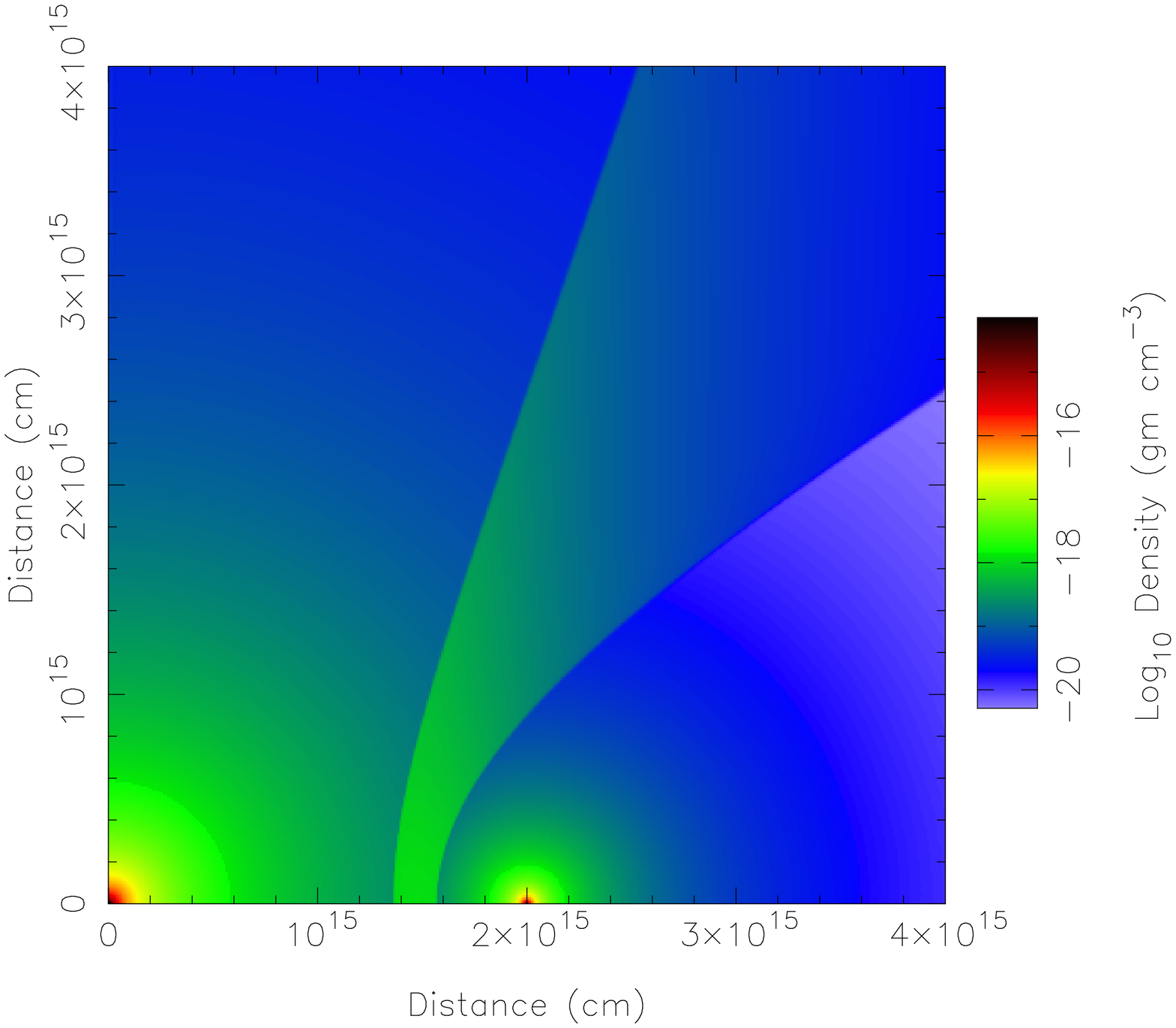}
 \includegraphics[width=8.5cm ]{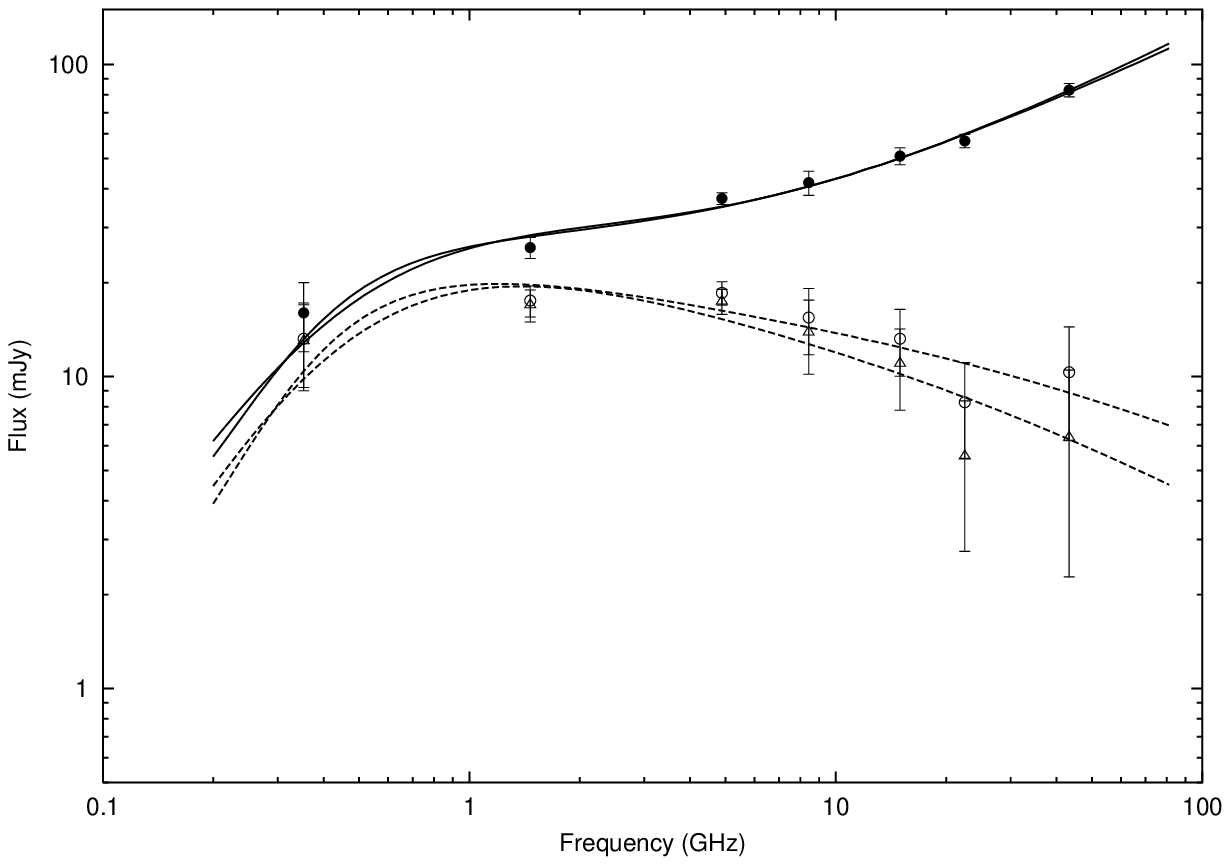}
 \caption{Left: Density distribution of a model CWB. The WR star is at
 (0,0) and the O star companion at (0,$2\times10^{15}$)~cm. Right:
 Models of the radio spectrum of WR147. The total (solid) and
 synchrotron (dashed) emission are shown for two possible
 models.\label{fig1}}
\end{figure}

To date, models of the radio emission from these systems have been
based largely on highly simplified models. As a first step toward more
realistic models, we have developed models of the radio emission based
on 2D axis-symmetric hydrodynamical models of the stellar winds and
the WCR. The temperature and density on the hydrodynamical grid
(Fig.~\ref{fig1}) are used to calculate the free-free emission and
absorption from each grid cell. We assume the electrons are
accelerated at the shocks bounding the WCR by DSA, and then advected
out of the system in the post-shock flow. At the shocks, the energy
distribution of the relativistic electrons is specified by a power-law
i.e. $n(\gamma)\propto\gamma^{-p}$.  This spectrum evolves away from a
simple power law as the electrons cool via several processes,
including inverse-Compton and Coulombic cooling. The models take all
these effects into account, including the impact of Razin effect.  The
synchrotron emission and self-absorption from each cell within the WCR
are then calculated from the local energy distribution of non-thermal
electrons. For more details see Dougherty et al. (2003) and Pittard et
al. (2005).

We have applied these models to observations of WR147
(Fig.~\ref{fig1}) and WR146 (see O'Connor et al., these proceedings)
with some success.  The models of WR147 are not as well constrained by
the available data as those from WR146. O'Connor et al. (these
proceedings) point out that models of WR146 would be much better if
the spectra steepened at high frequencies, and if the WCR emission was
less extended. We continue to investigate these issues.

\section{Observations and models of WR140}
The development of more realistic models of the radio emission from
CWBs was motivated, in part, by the difficulties earlier work had with
modelling the radio variations observed in WR140 (Williams et al., 1990; White \& Becker, 1995). In spite of the wealth
of multi-frequency observations of WR140, a number of critical system
parameters are either unknown (e.g. system inclination), or weakly
constrained (e.g. distance and luminosity class of the companion). Recent
radio observations now provide these key constraints.

\subsection{Modelling constraints from observations}
\begin{figure}
 \centering
 \includegraphics[width=8.4cm]{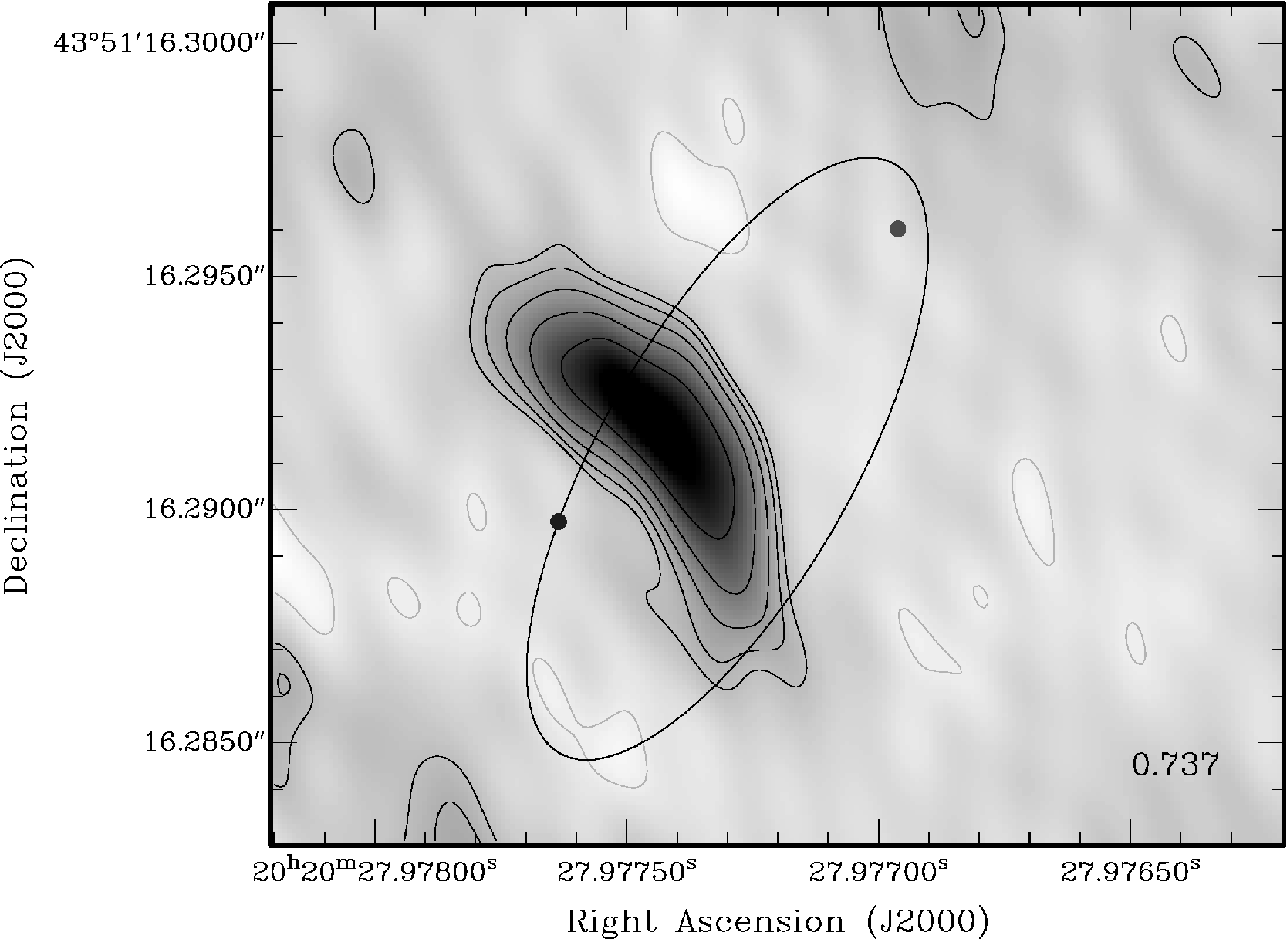}
 \includegraphics[width=8.4cm]{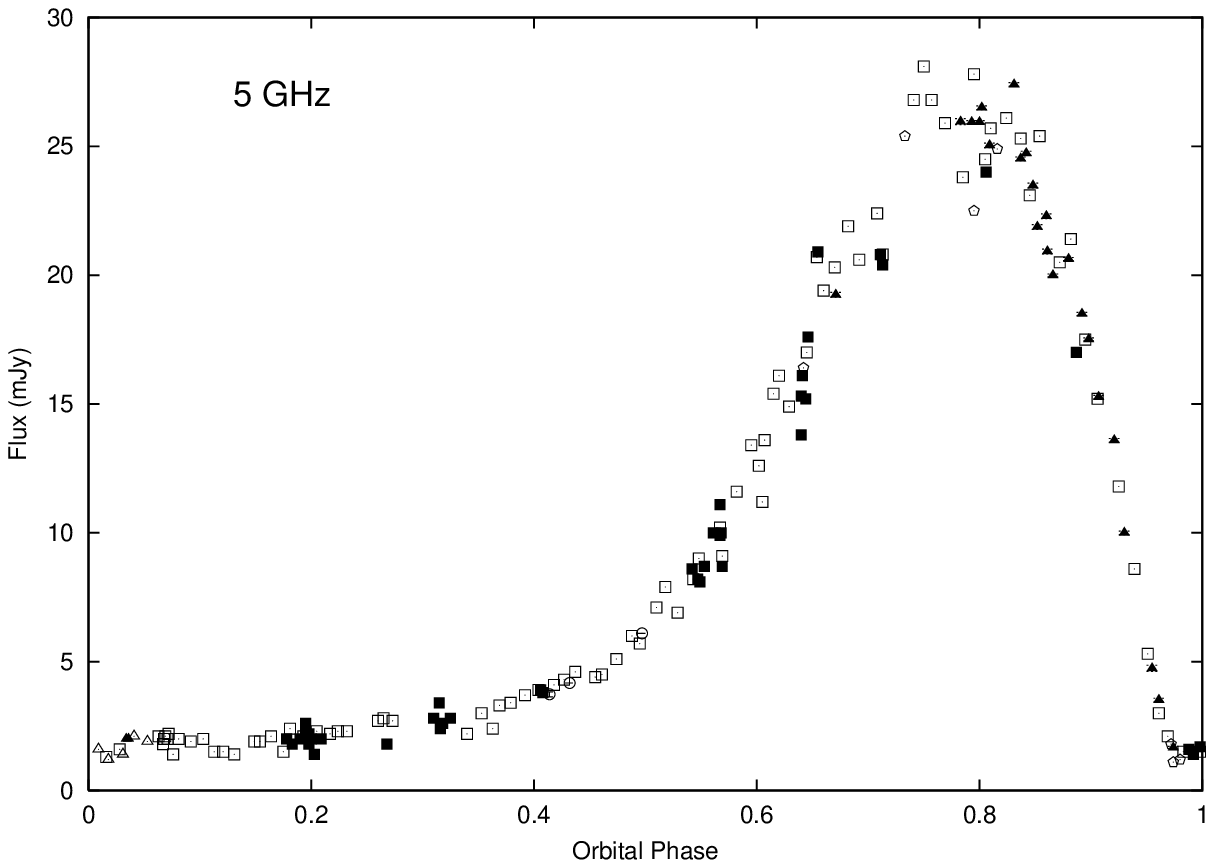}
 \caption{Radio observations of WR\thinspace140. Left: 8.4-GHz VLBA
 observation at orbital phase 0.737, with the deduced orbit
 superimposed. Right: 5-GHz flux as a function of orbital phase from
 orbit cycles between 1978-1985 (pentagons), 1985-1993 (squares),
 1993-2001 (triangles), and the current cycle 2000-2007
 (circles). Open symbols are from the VLA, and solid symbols from the
 WSRT.\label{fig2}}
\end{figure}

A 24-epoch campaign of VLBA observations of WR140 was carried out
between orbital phase 0.7 and 0.9. An arc of emission is observed,
resembling the bow-shaped morphology expected for the WCR
(Fig.~\ref{fig2}). This arc rotates from ``pointing'' NW to W as the
orbit progresses (Fig.~\ref{fig3}) which, in conjunction with the
observed separation and position angle of the two stellar components
at orbital phase 0.3 (Monnier et al. 2004), leads to a solution for the
orbit inclination of $122\pm5^\circ$ and the orbit semi-major axis of
$9.0\pm0.5$~mas. Using the $a\sin i$ derived by Marchenko et
al. (2003), we can derive a distance of $1.85\pm0.16$~kpc to
WR\thinspace140. This represents the first distance derived for CWB
systems {\em independent} of stellar parameters, and together with the
optical luminosity of the system implies the O star is a supergiant.
In addition, total flux measurements from the VLA (Fig.~\ref{fig2})
show that the radio variations from WR\thinspace140 are very closely
the same from one orbit to the next, pointing strongly toward
emission, absorption and cooling mechanisms that are controlled
largely by the orbital motion (Dougherty et al. 2005).
\subsection{Modelling the radio emission from WR140}
\begin{figure}
  \centering
  \includegraphics[width=8.4cm]{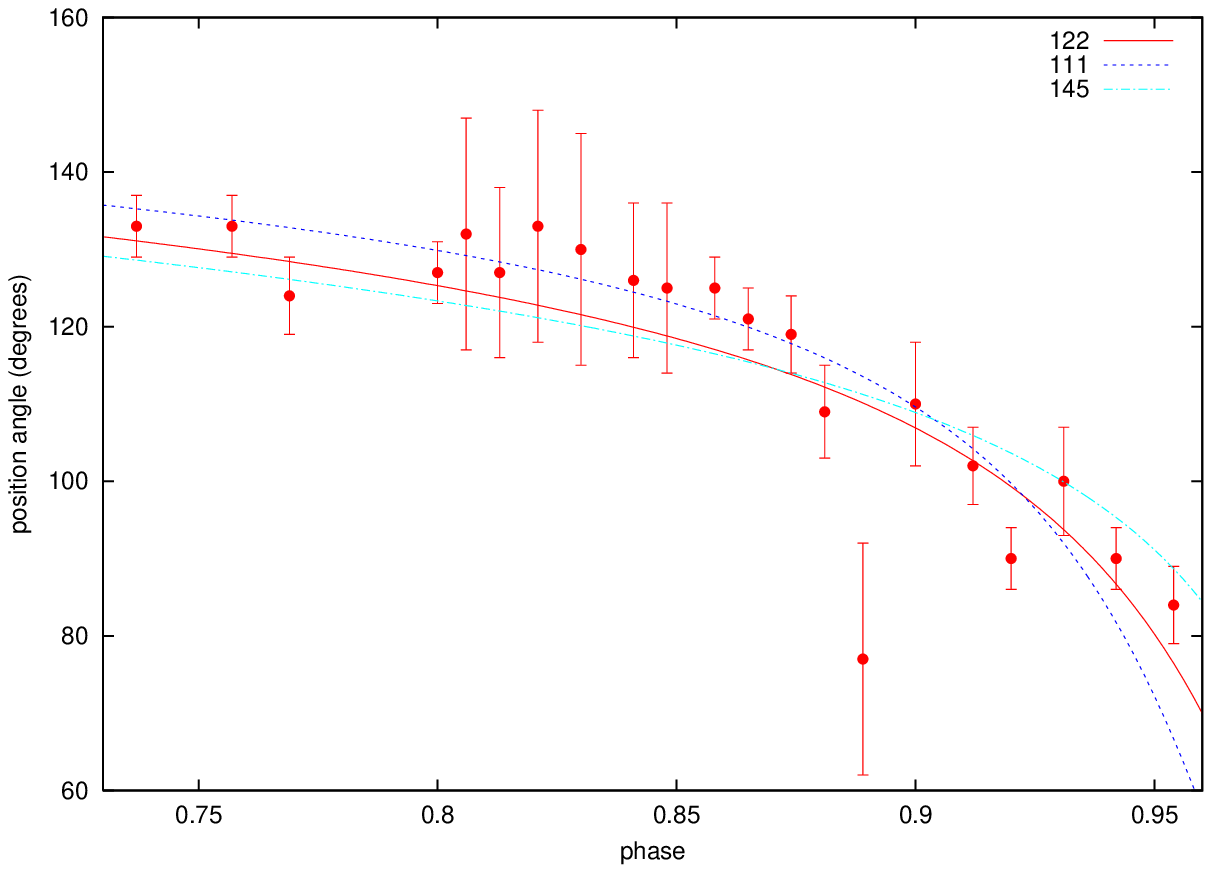}
  \includegraphics[width=8.4cm]{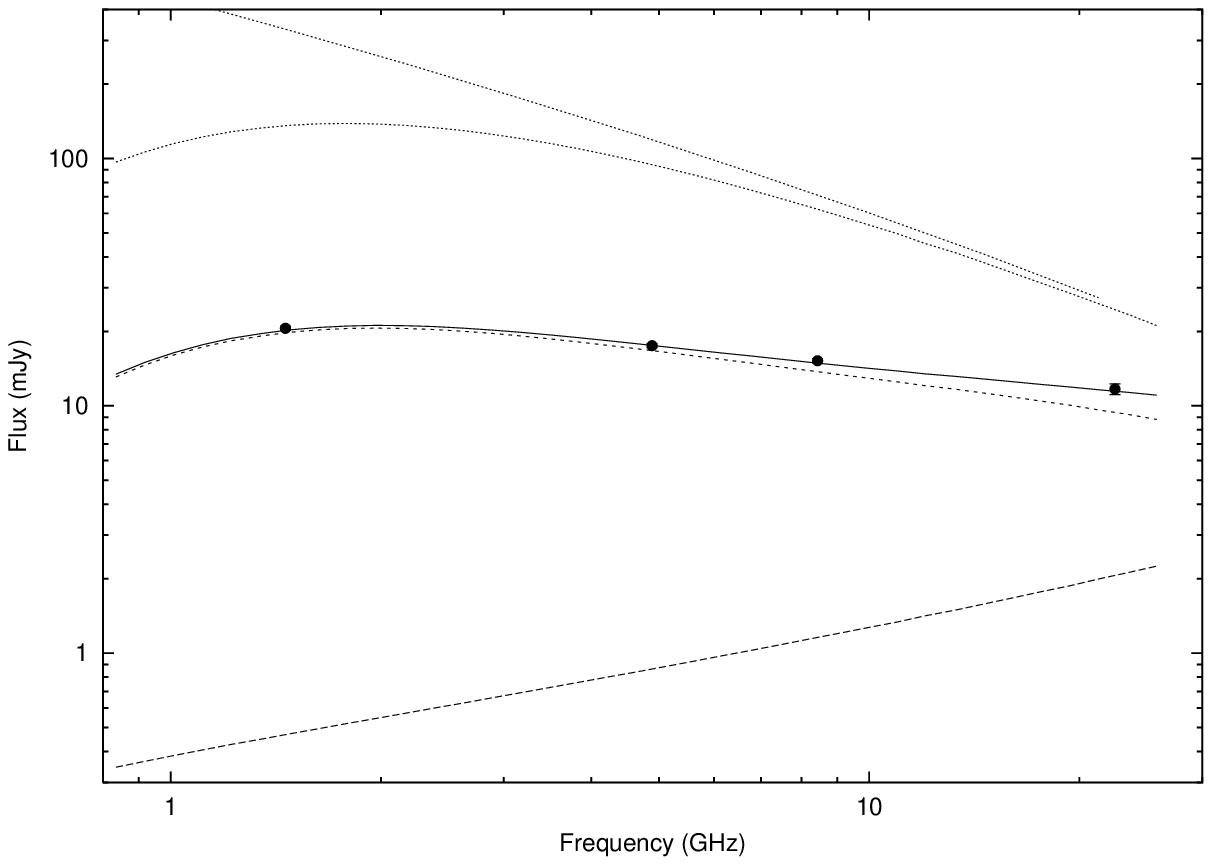}
  \caption{Left: The rotation of the WCR in WR140 as a function of
phase, deduced from VLBA observations. Curves derived from known
orbital parameters and three inclinations are shown, including the
best-fit of $122^\circ$. Right: The spectrum of WR140 at phase
0.9. The observations are the solid circles. Various emission
components from the model are shown - free-free (long-dashed),
synchrotron flux (short-dashed), intrinsic synchrotron flux (dotted),
and total flux (solid). The top curve shows the intrinsic synchrotron
spectrum without the Razin effect.
\label{fig3}}
\end{figure}

Using these new system parameters, we have applied our new
radiative transfer models to the spectrum of WR\thinspace140
in order to investigate the emission and absorption processes that
govern the radio variations. At orbital phase 0.9, an excellent fit to
the spectrum is possible (Fig.~\ref{fig3}). The free-free flux is
negligible compared to the synchrotron flux, which suffers a large
amount of free-free absorption by the unshocked O-star wind, as
anticipated at this orbital phase since the bulk of the WCR is
``hidden'' behind the photospheric radius of the O-star wind. The low
frequency turnover in this model is due to the Razin effect. Similar
fits can be determined for the spectrum at phase 0.8. However, we have
difficulty at earlier orbital phases ($\sim0.4$) if the low frequency
turnover is the result of the Razin effect, largely due to the low
value of magnetic field strength that is required, and the
commensurate extremely high acceleration efficiency that is
implied. We are continuing to explore these models.

The manner in which the non-thermal radio emission is calculated gives
both the spatial distribution of the magnetic field strength, the
intrinsic radio synchrotron luminosity, and the population and
distribution of non-thermal electrons. These can provide a robust
estimate of the high energy emission from CWBs, which is very relevant
given the current and planned high-energy observatories (see Pittard,
these proceedings).

%
%
\section*{Acknowledgments}
This work was supported by the National Research Council of Canada,
and the University of Prince Edward Island Co-op programme. JMP is
supported by a University Research Fellowship of the Royal Society,
and SMD acknowledges gratefully funding from the School of Physics and
Astronomy during a visit to Leeds.
%
%
 
\beginrefer
\refer Dougherty S.~M., Beasley A.~J., Claussen M.~J., \& Zauderer B.~A.,
  Bolingbroke~N.~J. 2005, \apj, 623, 447

\refer Dougherty S.~M., Pittard J.~M., Kasian L., Coker R.~F., Williams
  P.~M., \& Lloyd H.~M. 2003, \aap, 409, 217

\refer Dougherty S.~M. \& Williams P.~M. 2000, MNRAS, 319, 1005

\refer Dougherty S.~M., Williams P.~M., \& Pollacco D.~L. 2000, \mnras, 316,
  143

\refer Dougherty S.~M., Williams P.~M., van der Hucht K.~A., Bode M.~F.,
  \& Davis R.~J. 1996, \mnras, 280, 963

\refer Marchenko S.~V. et al. 2003, ApJ, 596, 1295

\refer Monnier J.~D. et al. 2004, ApJL, 602, L57

\refer Pittard,J.~M., Dougherty S.~M., Coker R., O'Connor E.P.,
  \& Bolingbroke N.~J. 2005, \aap, submitted

\refer White R.~L., Becker R.~H. 1995, ApJ, 451, 352

\refer Williams, P.M., Dougherty S.M., Davis R.J., van der Hucht K.A., Bode M.F., Setia Gunawan D.Y.A. 1997, \mnras, 289, 10

\refer Williams P.~M., van der Hucht K.~A., Pollock A.~M.~T.,
Florkowski D.~R., van der Woerd H., Wamsteker W.~M. 1990, MNRAS, 243,
662 

\refer van Loo S., Runacres M.C., \& Blomme R. 2005, \aap, submitted.

\endrefer

\end{document}